\documentstyle[aps,twocolumn]{revtex}
\input psfig
\begin{document}

\def\om{\omega}
\def\Om{\Omega}
\def\l{\lambda}
\def\eps{\varepsilon}
\def\siml{\hbox{\kern.1em \lower.6ex \hbox{$\sim$} \kern-1.12em
          \raise.6ex \hbox{$<$} \kern.1em }}
\def\simg{\hbox{\kern.1em \lower.6ex \hbox{$\sim$} \kern-1.12em
          \raise.6ex \hbox{$>$} \kern.1em }}

\title{Super Shell Structure of the Magnetic Susceptibility }
\author{S.Frauendorf}
\address { Institute for Nuclear and Hadronic Physics,
Research Center Rossendorf, PB 51 01 19,  01314 Dresden, Germany }
%\address{Cyclotron Institute, Texas A \& M University, College
%Station, Texas 77843-3366, USA
\author{  V.M.Kolomietz, A.G.Magner and A.I.Sanzhur} 
\address{Institute for Nuclear Research, 252028 Prospekt Nauki 47,
Kiev--28, Ukraine}
\maketitle

\begin{abstract}
The magnetic susceptibility of electrons confined to a spherical cavity
or a circular billiard shows slow oscillations as a function of the number 
of electrons, which are a new manifestation of the Super Shell Structure 
found in the free energy of metal clusters. The relationship of the 
oscillations of the two different quantities is analyzed by means of  
semiclassical calculations, which are in quantitative agreement with 
quantal results.
  The oscillations should be observable for ensembles of 
circular ballistic quantum dots and metal clusters.

\bigskip\noindent
PACS numbers: 05.45.+b, 02.50.+s,03.65.-w,03.65.Ge
\end{abstract}

\bigskip

\section{Introduction}
The confinement of independent Fermions in 2 or 3 dimensions (2D or 3D) 
leads to a bunching of the  single particle levels, 
if the mean free path of the Fermions is large compared to the 
size of the system. This is known as shell
structure (SS) and leads to oscillations of the total energy as
function of particle number $N$ around a smoothly changing background.
The oscillating part, referred to as shell energy, has minima at the
so called magic numbers, which  have been known for a long time for
nuclei. More recently they have been observed in the
abundance spectra of alkali metal clusters (c.f. \cite{deHeer} and
the original work cited therein), representing minima of the free energy
\cite{hansen}.
Later on, it has been found \cite{pedersen}, that the amplitude of these  
shell oscillations  is modulated by a slow oscillation. This so called 
Super Shell Structure (SSS) had been predicted theoretically 
\cite{bablo,nishioka}. In this paper we will show that the magnetic 
susceptibility follows a similar SSS pattern. Using the 
Strutinsky's shell correction method \cite{strut1,strut2,frapasrot}
and semiclassical Periodic Orbit 
Theory (POT), we will trace  the SSS of the susceptibility and 
free energy back to the same interference pattern between electrons
on classical periodic orbits.  

The consequences of SS  for the  magnetic
susceptibility of the confined electron gas have been
discussed in refs. \cite{levy,ruitenbeck,oppen,ullmo,richter} and earlier
references cited therein. 
The susceptibility of a confined ballistic  2D  electron gas 
can be measured for
 large  ensemble of  quantum dots  on a \mbox{AlGaAs} - GaAs 
semi conductor hetero  structure \cite{levy}.
For this type of experiments it is claimed  \cite{levy,oppen,ullmo,richter}
 that the shell
oscillations as a function of the electron number  $N$ 
are averaged out by the fluctuations of the
size and the shape of the individual dots.
 The only Quantum Size Effect (QSE) expected to
survive is a paramagnetic enhancement of the
susceptibility, which changes smoothly with $N$.  
In this paper we will
argue that such experiments
should permit to resolve the   slow oscillations reflecting the SSS.
We will also discuss the experimental possibilities to detect SSS of
the magnetic susceptibility of metallic clusters.   

\section{Susceptibility of electrons in a spherical cavity}

Choosing $z$ as the direction of the magnetic field $H$, the 
orbital part of the electronic
Hamiltonian is \cite{kittel}
\begin{equation}
{\cal H}={\cal H}_0 + \om L_z  + {M \om^2 \over 2} \left(x^2+y^2\right)
                                                              \label{hamilt},
\end{equation}
where ${\cal H}_0$ is the Hamiltonian at the zero magnetic field,
consisting of kinetic energy and the confining potential, which is
assumed to be circular or spherical. The operator $L_z$ is the angular
momentum projection on  the $z$-axis and $M$ the effective electron mass.
We use the Larmor frequency  $\om=\mu_B H/\hbar$ as the unit of the 
magnetic field $H$ in order to stress the analogy to the case of a 
system rotating with the angular velocity $-\om$.
%The electronic spin is not considered. The justification will be given 
%below.

Up to third order of  perturbation theory in $\om$, the thermodynamical
potential $\Om(T,\l,\om)$ as function of the temperature $T$ and chemical
potential $\l$ is
\begin{eqnarray}
& \Om(T,\l,\om) = -T \sum_{\nu} \ln \bigg[ 1+ & \nonumber \\
& \exp\left({\displaystyle {\l\!-\!\eps_{\nu}\!-\!\hbar\om 
m_{\nu}\!-\! M\om^2\langle x^2\!+\!y^2\rangle_{\nu}/2 \over T}}
\right)\bigg], & 
                                                             \label{ompoten}
\end{eqnarray}
where $\eps_{\nu}$, $m_{\nu}$ and $\langle\ldots\rangle_\nu$
are the energy, the angular momentum projection and the expectation
value with the unperturbed electron state $\nu$. In our units, the
zero field susceptibility is a moment of inertia. For the
grand canonical ensemble it reads
\begin{eqnarray}
\theta=-\left( {\partial^2 \Om \over \partial\om^2} 
\right)_{\om=0} = \theta_{cr}-\theta_{rig},
                                                                   \label{dchiom}\\
\theta_{cr}=\sum_{\nu} \left( \hbar m_{\nu} \right)^2 
{\partial n_{\nu} \over \partial\l}, 
                                                                                \label{thetacr}\\
\theta_{rig}=M\int d{\bf r}\,\rho({\bf r})
                                                             \left(x^2+y^2 \right),
                                                                                \label{inertrig}
\end{eqnarray}
where 
$n_{\nu}=(1+\exp[(\eps_{\nu}-\l)/T])^{-1}$ are the Fermi occupation
numbers and $\theta_{rig}$ is the moment of inertia of rigid rotation,
$\rho ({\bf r})$ being the particle density.

The shell structure of $\theta_{cr}$ has been analyzed for nuclear
rotation (where it is called cranking moment of inertia).  
Following the concepts of Strutinsky's shell correction method
\cite{strut1,strut2}, $\Om$, $\theta_{cr}$ and other
quantities are divided into a smooth part  and into an 
shell part (denoted by the 
subscript SH). The shell contribution to $\theta_{cr}$, which we call
$\theta_{SH}$, 
 represents the total QSE, 
because the shell part of $\theta_{rig}$
is negligible as compared to $\theta_{SH}$.
 The partition can either be done numerically
starting from the quantal electron levels \cite{strut1,strut2,frapasrot} 
or it can be based on semiclassical periodic orbit theory (POT).
In the latter case the shell terms read
\cite{kolmagstrut,magkolstrutsem,richter} 
\begin{eqnarray}
\left\{ {\theta_{SH}(T,\l) \atop \Om_{SH}(T,\l)} \right\} = 2
\sum_{\beta} \left\{ {a\,(\ell_{\beta})^2 \atop (\hbar/\tau_{\beta})
^2} \right\}  
A_{\beta}(\l)~\times      \nonumber \\
\sin \left( \frac{1}{\hbar} S_{\beta}(\l) + \nu_{\beta} \right)
{T\tau_{\beta}/\hbar \over \sinh(T\tau_{\beta}/\hbar)}~,
                                               \label{thetascl}
\end{eqnarray}
where $\Om_{SH}$ denotes the value at zero field and
$a=1$ or $\frac{1}{3}$ in 2D or 3D, respectively.
For the spherical cavity \cite{bablo} and the 
circular billiard \cite{ullmo,richter} the orbits $\beta(t,p)$
are defined by the number $t$ of the revolutions around the center and
the number $p$ of the corners,
\begin{eqnarray}
L_{\beta} = 2pR\sin\phi,~~\phi=\pi t/p, \\
S_{\beta}=\hbar k L_{\beta},~
\ell_{\beta}=\hbar kR \cos \phi,~
\tau_{\beta}=\frac{ML_{\beta}}{\hbar k}, \\
A_{\beta} = \frac{2MR^2}{\hbar^2}\frac{1}{\sqrt{kR}}
\frac{f_{\beta}\,(\sin{\phi})^{3/2}}{\sqrt{p\pi}},~~(p \ge 2t),~~~2D,
                                                                                                                             \label{amp2d} \\
A_{\beta} = \frac{2MR^2}{\hbar^2}\sqrt{kR} \sin(2\phi)
\sqrt{\sin{\phi} \over {p\pi}},~~(p > 2t),~~~3D,
                                                                                                                              \label{amp3d} \\
A_d= \frac{2MR^2}{\hbar^2}\frac{1}{p \pi} ~~(p=2t),~~~3D~\\
\nu_{\beta}=-{3\pi \over 2}p +{3\pi \over 4}, ~~~2D, \\
\nu_{\beta}=-{3\pi \over 2}p -(t-1)\pi-{\pi \over 4}, ~~~3D,
\end{eqnarray}
where we have introduced the length of the orbit $ L_{\beta}$,
the wave number $k=\sqrt{2M\l/\hbar^2}$ and 
$f_{\beta}=1$ 
for diameters and 2 for planar orbits.
The phases $\nu_{\beta}$, which 
are related to the Maslov indeces,
are not important in our discussion 
and are given in \cite{bablo,richter}.
 The POT level densities of a 
spherical cavity and a circular billiard in a magnetic field have been 
studied in refs. \cite{dotreim,tanaka}.

Since the energy to extract one electron from the confining potential is 
much higher than the temperature, it is important to use the canonical 
ensemble and define the susceptibility as the derivative 
$\theta= -(\partial^2 F/\partial \om^2)_{\om=0}$
of the free energy at fixed particle number $N$.
The importance of the fixed electron number for the magnetic properties
of 2D-structures has been pointed out previously \cite{ullmo,richter}.  
We adopt the approximation valid for large $N$, calculating
$F(T,N,\om)=\Om(T,\l,\om)+\l N$, where $\l$ is found from the
condition 
\begin{equation}
-\partial\Om(T,\l,\om) / \partial\l = N~.
\label{partnum}
\end{equation}
The zero field
susceptibility is given by the  expressions (\ref{dchiom}) or 
(\ref{thetascl}) taken at $ \l(N)$  fixed by eq.~(\ref{partnum}) at 
$\om=0$.
%the equation \mbox{$-\partial \Om(T,\l(N),\om=0) / \partial\l = N$}. 
We solve this equation numerically both for the semiclassical and the 
 quantum calculation. 
% because we 
%found that the susceptibility substantially deviates from the values   
%obtained by linearizing eq. (\ref{partnum})  as in  refs. \cite{ullmo,richter}.

\section{Spherical cavity}
 
Fig. 1 shows $\theta_{SH}(N,T)$ calculated by means of the numerical 
Strutinsky averaging procedure \cite{frapasrot} from the quantal levels 
in a spherical cavity. 
%with the radius $R=r_S N^{1/3}$.
%, the chemical potential
%$\l=\eps_F=50.1 eV(a_o/r_S)^2$ 
%We assume  id fahe effective mass $M =M_o$ ($M_o$ free electron mass) and  
%id fahe Wigner - Seitz radius $r_S=0.208~ nm$ ,
The parameters are  appropriate for sodium. 
As a unit we use the Landau diamagnetic susceptibility (LU) for the 
electron gas in the cavity, $|\theta_{L}|=0.2715 MNr_S^2$. Since all 
contributions to the bulk susceptibility are of the same order of 
magnitude \cite{kittel}, the figure shows directly the enhancement due 
to the QSE.

\begin{figure}[t]
\vskip-1cm
\hspace{-1cm}
%\hspace{-3cm}
%\mbox{\psfig{file=sphere.ps,width=18cm}}
\mbox{\psfig{file=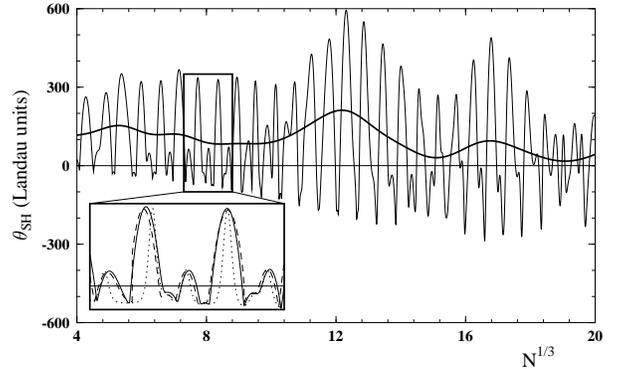,width=9.5cm}}
\vspace{-5cm}

\caption{
Shell contribution $\theta_{SH}$ of the magnetic susceptibility of
$N$ electrons in a spherical cavity. The effective mass is $M=M_e$, 
the radius  $R=r_s N^{1/3}$ with $r_s =0.208~ nm$,
the Fermi energy $\eps_F=3.24~ eV$ and the temperature  
$T=0.005\eps_F=170~ K$. 
The thin line shows the results without averaging
over $N^{1/3}$ and the fat ones after averaging with a Gaussian 
of width $\Delta N^{1/3}= 0.75$. Inset: solid and dashed lines
show the quantum and semiclassical calculations for the canonical ensemble,
respectively. Dots present the grand canonical quantum result (fixed
$\l=\eps_F$).}
\end{figure}

The susceptibility $\theta_{SH}$ oscillates with the period of the 
shells. In addition its amplitude is modulated with  a slow  oscillation,
which is the ``Super Shell Structure'' (SSS), 
first  noticed for the level density \cite{bablo,nishioka}. It was found
experimentally in the abundances of Na clusters \cite{pedersen},
which are determined by $F_{SH}$\cite{hansen}, showing also the SSS pattern
(cf. Fig. 2). 

We  have also evaluated $\theta_{SH}$ by means of the POT sums
(\ref{thetascl}).
As seen in the inset of Fig. 1, the quantal values of $\theta_{SH}$ 
agree very well with ones obtained from the semiclassical expression.
% The quantitative agreement is only obtained
% with the {\it exact} chemical potential $\lambda(N)$ calculated by solving
%eq. (\ref{partnum}) numerically. 
 Semiclassics  permits 
a simple interpretation of  the SSS. The shortest PO's
enclosing magnetic flux are the 
triangle and square. The temperature factor, 
%in eq. (\ref{thetascl}), 
%$ 1/\sinh{(-T\tau_{\beta}/\hbar)}\approx \exp{(-L_\beta/\Lambda_T)}/2$, 
%where $\Lambda_T=TM/\hbar k$ 
$1/\sinh(T\tau_\beta/\hbar)\approx 
2 \exp(-L_\beta/\Lambda_T)$~, where $\Lambda_T=\hbar^2 k/TM$~
is the characteristic
 ``temperature  length'',
damps the  longer orbits.
%in the POT sum (\ref{thetascl}) but not in $\lambda(N)$ itself.
The beat pattern results from the 
superposition of the two leading terms, the triangle and the square. 
The basic oscillation has a 
period given by $k(L_\triangle+L_\Box)/2$ and the  beat oscillates   
with  $k(L_\Box-L_\triangle)$.
Fig. 3 shows a calculation that takes into account only the triangle
and the square.  Comparing with the full calculation,
the influence of the longer orbits is seen. Although they make the peaks
 higher the   beat pattern is not much changed up to $N^{1/3}\sim 16$.
The upper and lower envelops of the full calculation are not very
 different from the ones of the truncated calculation. This demonstrates 
that the beat pattern is basically generated by the triangle 
and the square. For $N^{1/3} \simg 16$,  
the full calculation has a beat minimum
where the truncated one has a maximum, indicating that the interference 
with the longer orbits becomes important.

\begin{figure}[t]
\vskip-1cm
\hspace{-1cm}
%\hspace{-3cm}
%\mbox{\psfig{file=f_sh-sph.ps,width=18cm}}
\mbox{\psfig{file=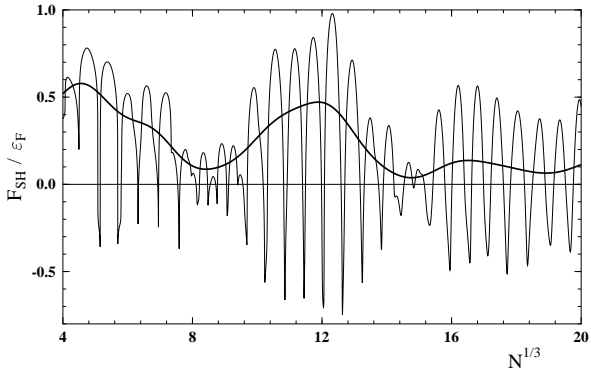,width=9.5cm}}
\vspace{-5cm}
\caption{Shell contribution $F_{SH}$ to the free energy of
$N$ electrons in a spherical cavity. The calculation and the
 conventions are identical with Fig. 1.}
\end{figure}

\section{Relation to the shell structure in the free energy}

The relation between the shell contributions to the zero field
free energy, $F_{SH}$, and to the susceptibility, $\theta_{SH}$,
is understood by comparing the two POT sums (\ref{thetascl}). The terms are
identical up to the factor 
\begin{equation}\label{scaling}
\Upsilon_\beta=a(l_\beta \tau_\beta/\hbar)^2=
\hbar^{-2}M^2R^4ap^2\sin^2 (2\phi)
\end{equation}
in $\theta_{SH}$, which suppresses the orbits
$l_\beta=0$ and gives the long orbits a higher weight.  
If only few orbits with  similar values of $\Upsilon_\beta \approx \Upsilon$
contribute, $\theta_{SH}\approx \Upsilon F_{SH}$. 
The simple scaling is also expected 
to hold for  shapes not too different from the sphere and can be used to 
relate the SS in the susceptibility and free energy.

In the 3D case  the diameter orbit  is suppressed
by a factor $1/\sqrt{kR}$   as compared to the planar orbits,
 because it has a lower degeneracy.
 This makes its contribution to $F_{SH}$  insignificant for large $N$. 
The two sums $F_{SH}$ and 
$\theta_{SH}$ become similar,
because the leading terms  are the triangle and square. 
 In fact, for $T> 0.02 \eps_F$
we find $\theta_{SH}\approx \Upsilon F_{SH}$ with 
$p=3-4$ in (\ref{scaling}) 
(where $\eps_F$ is the Fermi Gas energy). For the lower temperature
$T=0.005\eps_F$, shown in Figs. 1 and 2, we find a ratio $\Upsilon$
with $p=4$ at the SSS maxima ($N^{\frac{1}{3}}=12$ and 17).
 The SSS minima are less pronounced
 for  $\theta_{SH}$ than for $F_{SH}$.
There the triangle and square cancel each other.
The main contribution comes from longer orbits,
which are much more important for $\theta_{SH}$ than for $F_{SH}$,
 preventing $\theta_{SH}$ from becoming as small as $F_{SH}$. 
This interpretation is also supported by Fig. 3. The difference between the 
full  and truncated calculations, which represents the contribution of the
longer orbits, is just an up-shift of the upper envelop .

\begin{figure}[t]
\vskip-1cm
\hspace{-1cm}
%\hspace{-3cm}
%\mbox{\psfig{file=t34sph.ps,width=18cm}}
\mbox{\psfig{file=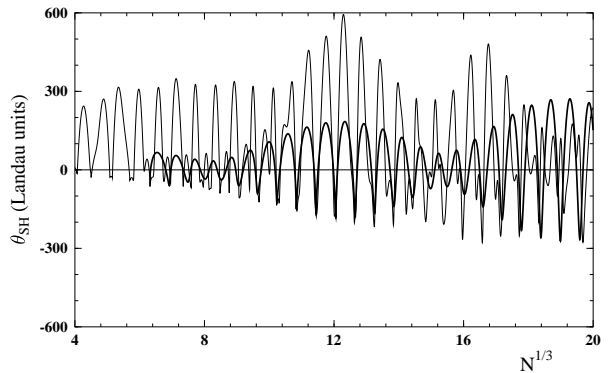,width=9.5cm}}
\vspace{-5cm}

\caption{
Shell contribution $\theta_{SH}$ of the magnetic susceptibility of
$N$ electrons in a spherical cavity.
The thin line shows the same as fig. 1, the thick line corresponds 
to a calculation where only the triangular and square orbits in the 
POT sums (\protect\ref{thetascl}) are taken into account. For 
$N^{1/3} <6$ the level density is not positive definite in this approximation
and the solution of eq. (\protect\ref{partnum}) is not possible.}
\end{figure}

The SSS is clearly developed for the
temperature of $T=0.005\eps_F$,
shown in  Fig. 1.
For $T=0.0005\eps_F$ we find very pronounced SS with an amplitude of a few 
thousand LU, but little SSS.
The averaged susceptibility stays around 1500  LU
with small peaks (400  LU high) at $N^{1/3}=7$ and 13.
Due to a weak temperature damping in the POT sum
many orbits contribute to the sum, destroying the simple beat pattern.
This is at variance with  a distinct SSS pattern 
in $F_{SH}$  persisting to $T=0$ \cite{bablo,nishioka}. The reason is
the factor $1/\Upsilon_\beta$ which  suppresses the orbits with high $p$. 
 For  $T=0.05\eps_F$  there is SSS, but  the
amplitude of the basic shell oscillations remains below 5 LU for $N<100$,
becoming small compared with the LU  for larger $N$. Hence, the 
intermediate temperature seems to be optimal for the observation of  the SSS. 

\section{Canonical ensemble}

Experimentally, the average susceptibility of an ensemble of
clusters  with a distribution in $N$ will be measured. Let us consider
a Gaussian distribution  in $N^{1/3}$ 
of width  $\Delta (N^{1/3})=0.75$, which corresponds to about one oscillation.
As shown in Fig.1, {\em averaging the 
susceptibility damps out only the basic
shell oscillations, whereas the SSS remains as a modulation of the strongly
paramagnetic susceptibility.}
Hence, it is expected that the SSS can be observed with 
a moderate 
mass selection of the clusters that would not allow to resolve
the basic SS ($\Delta N/N=0.23$ for $N=1000$ in Fig. 1).

The QSE of the susceptibility survives the averaging
only for the canonical ensemble \cite{levy,oppen,ullmo,richter}.
 The inset of Fig. 1 compares 
the canonical with grand canonical ensemble  ($\l=\eps_F$).
Though the  positions of the extrema are similar, the shape of peaks
is rather  different. 
For the grand canonical ensemble
the negative and positive values are equally probable and, as the result,
averaging with respect to $N$ quenches $\theta_{SH}$.
For the canonical ensemble, 
the positive values are more frequent and  the QSE survives
the averaging with respect to $N$.
 The preference of $\theta_{SH}>0$ for the canonical
 ensemble is
evident from  POT sum  (\ref{thetascl}): 
The shell correction to the level density  $g_{SH}$, which is
given  by setting the factor $\{...\}=1$, oscillates in phase with 
$\theta_{SH}$, i. e.    more  
$N$ values correspond to a paramagnetic than to a 
diamagnetic susceptibility and the average with respect to the particle
number is positive.

Let us discuss this correlation in more detail, because it is crucial 
for the measurability of the SSS.  
Fig. 3 illustrates that the typical inverted parabolas 
appear already for the lowest orbits.
It is sufficient to consider only one term in (\ref{thetascl}), say the 
triangle. For the grand canonical ensemble, the wave number $k=k_F$ is
 constant ($k_F$ is the Fermi Gas value).  The orbit length is
 $ L_\triangle \propto N^{1/3}$. The susceptibility
is proportional to $\sin (k_FL_\triangle)$ (for simplicity the phase 
$\nu_\triangle$ is left away), 
which averages to zero.
The particle number expectation value as given by (\ref{partnum}) 
contains a  term proportional to
$ -\cos (k_FL_\triangle)$ that makes it oscillating around 
$N$. Thus, for the canonical ensemble, the wave number $k=k_F+\delta k$ 
cannot be  constant. It must contain an oscillating
term $\delta k$  in order to satisfy eq. (\ref{partnum}).  
In order to understand qualitatively its consequences for the susceptibility
we use an argument from refs. \cite{ullmo,richter}. 
 Retaining only the   
linear order of  $\delta k $,  eq. (\ref{partnum})  gives 
$\delta k \propto \cos (k_FL_\triangle)$.
The susceptibility is  proportional to 
\mbox{$\sin ((k_F+\delta k)L_\triangle)\approx \sin (k_FL_\triangle)+
\delta kL_\triangle \cos (k_FL_\triangle) $}. The second term is
proportional to  $\cos (k_FL_\triangle)^2 $ which averages to 1/2. 
Hence, the total averaged susceptibility is positive.
Since the interference between the triangular and quadratic
orbits in both $\theta_{SH}$ and $g_{SH}$  is about the same,
the SSS modulates paramagnetic term also after averaging.
The oscillations of $\delta k$  narrow 
the minima and broaden the maxima of the susceptibility. 
In our calculations, $\delta k $
is treated exactly. As seen in Figs. 1 and 3, due to the higher 
orders in $\delta k $ the minima are narrowed to   
cusps and the maxima take the shape of parabolas. 

\section{Real clusters}

Real clusters deviate from the perfect spherical cavity:
The surface has a finite thickness of the order of the screening length.
 The discrete
ionic back ground implies a certain surface roughness of the
 order of the interatomic distance. There may be
impurities or other imperfections distributed over the volume. 

The SS in a spherical potential with a realistic surface thickness
(the one of sodium) has been studied in ref. \cite{nishioka}.
The SSS in the binding energies is clearly developed. The 
beat minima are somwhat shifted as compared to the cavity. The shift
has been be traced back  to small changes of the action of the POT due
to the modified ``reflection'' by the finite potential at the surface.
Since  the expressions (\ref{thetascl}) hold also for the more realistic
potentials if the appropriate action $S$ is inserted, a similar shift of the
SSS pattern can be infered for the susceptibility.       

The consequences of the surface roughness 
for the shell structure of the ground state energy
have been studied in ref. \cite{rough}. The energy is given by 
the POT sum (\ref{thetascl}), where each term contains an additional 
damping factor $\chi^p$. Assuming that the rough surface is randomly
displaced relative to the ideal one, 
the damping factor is  
\begin{equation}\label{rough}
\chi^p=\exp{(-2p(\sigma k\sin{\phi})^2)},
\end{equation}  
where  a Gaussian displacement distribution with the width $\sigma \sim r_S$
is used.
It has a simple interpretation. The number of reflection on the  
surface is $p$.
For each reflection  the rough surface scatters
a certain fraction of the particles   away  from the POT.  
The arguments of ref. \cite{rough} can be immediately  taken over to
the zero field   susceptibility. The damping factor arises from the 
reflections on the irregular surface. A weak magnetic field does not
change the reflections on the surface
\footnote{The difference beween the case with a weak  and without a field
is the  small curvature of the trajectory
 between the reflection points which barely changes the angles of the 
trajectory with the surface.} and the  damping factor (\ref{rough})
appears also in the susceptibility.  
Hence, the long orbits (large number of reflections $p$) 
 are  strongly suppressed by the surface roughness. In ref. \cite{rough}
it is shown that a roughness $\sigma = 0.2 r_S=0.38/k_F$ 
reduces the amplitude of the 
SS to about 1/2 of the one of the ideal cavity. The SSS is found to be
 nearly the same
as for the ideal cavity. For the susceptibility one expects that the 
surface roughness strongly reduces the contributions of the long orbits,
such that the SSS pattern of Fig. 1  approaches  the one of Fig. 3 with a 
reduced amplitude. In particular, for very low temperature the surface
roughness is expected to enhance the SSS, because it efficiently damps the
long orbits.     

Impurities or other imperfections
that are homogeneously distributed over the volume will also 
scatter the particles away from the POT. The situation 
is analogous to the propagation of a wave in an absorbing medium, 
which has been considered in ref. \cite{bablo}. The scattering results in a 
damping factor of the form $\exp{(-L_\beta/\Lambda_I)}$, where
$1/\Lambda_I$ measures the amount of scattering per unit length 
and corresponds to the mean free path of the electrons due to the
 imperfections.  
This kind of damping  is equivalent with an 
increase of the temperature, because
the temperature damping factor in (\ref{thetascl})
$1/ 2\sinh(T\tau_{\beta}/\hbar)\approx \exp{(-L_\beta /\Lambda_T)}$.
The consequences of a temperature
 increase are discussed above.

\section{Measurements of the susceptibility of metal clusters}
 
In Ref. \cite{kimura} the susceptibility has been measured for
three  ensembles of clusters with  the average size of $N^{1/3}\approx 4$, 
2 and 1.5 and a large  spread in size.
A paramagnetic enhancement of 5, 2 and 1.5 is found, respectively.
The decrease of the enhancement with $N$, which  ref. \cite{kimura}
mentions as an unexplained phenomenon, can be seen in Fig. 1.  
To identify the SSS, experiments with more points in $N$ and a
mass resolution better than 30\% are needed. 
An alternative experiment would be the measurement of the deflection
of a cold cluster beam in an inhomogeneous magnetic field, which
provides directly the susceptibility. To reach the necessary sensitivity
of such a  Stern-Gerlach apparatus seems to be possible \cite{beam}.
It is favourable for this kind of experiment that only a moderate mass
selection is needed to observe SSS, what permits larger intensities.

Measuring the  susceptibility could shed new light on the
electronic structure  of small metal clusters.
Solid icosahedral shapes
have been suggested for   sodium clusters  with $T \sim 200 K$ \cite{martin}.
Our sphere model
should be a rough first approximation for $N < 1000$. In this range   
the shell energies $E_{SH}$ for the spherical and icosahedral cavity
 are similar \cite{pavlov} and, as discussed above, 
 the same can be expected for the susceptibilities.
A more pronounced paramagnetic 
SSS is expected if shape of the cold clusters
comes close to a rough sphere. 
The picture changes,  
 if the clusters were 
liquid or would keep keep the same shape as in the liquid 
state when  freezing at low $T$.
Then magic clusters  are spherical and strongly
diamagnetic, whereas the non magical ones are deformed and weakly diamagnetic 
\cite{frapasrei}. Thus, the averaged susceptibility would be diamagnetic, 
 still showing a SSS.

\section{Half sphere}

Fig. 4 shows the susceptibility for a half sphere with 
$R=r_S(N/2)^{1/3}$.
The results are very similar to the full sphere. The main difference
is a shift of the basic shell oscillations by half a period. 
The averaged susceptibility is almost indistinguishable from Fig. 1.
 Cluster deposited
on an insulating surface may take shapes close to half spheres 
\cite{surface1,surface2,surface3}.

\begin{figure}[t]
\vskip-1cm
\hspace{-1cm}
%\hspace{-3cm}
%\mbox{\psfig{file=halfsph.ps,width=18cm}}
\mbox{\psfig{file=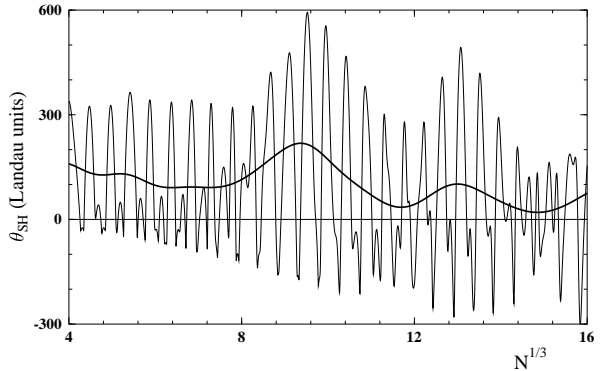,width=9.5cm}}
\vspace{-5cm}
\caption{Shell contribution of the magnetic susceptibility
$N$ electrons in a half spherical cavity.  The line
 conventions and calculation (except $R=r_S(N/2)^{1/3}$)
are identical with Fig. 1.}
\end{figure}

\section{Circular quantum dots}

The  susceptibility of the 2D  electron gas confined to a circular 
potential well with $R= 500~ nm$  is shown in Fig. 5.  An effective mass of 
$M=0.067M_{e}$ \cite{weiss}, appropriate for GaAs, is assumed.
The LU 
is $|\theta_L|=MR^2/3$.  The 2D case is similar
to the 3D - case, the main difference consisting in an increase of  
the QSE with $N$ (The values in Fig. 5 are divided by $N$). 
At the temperature
$T= 5\hbar^2/2MR^2 \approx 0.13K$ a distinct 
 SSS is seen. For  $T= 0.5\hbar^2/2MR^2$  the averaged
 susceptibility $\sim 4/N$ LU showing  shallow oscillations 
with an  amplitude of $\sim 1/N$ LU, which loosely correlate with the SSS
in Fig. 5. A SSS beat pattern for a circular dot has been first calculated in
\cite{richter}.  However, it is 
argued there and also  in \cite{oppen,ullmo} that when averaging over the
 ensemble of dots used as experimental probe, uncertainties in the shape 
and size will completely 
wipe out  the shell structure, the only remaining QSE being a paramagnetic
enhancement that varies smoothly with $N$.   In contrast, 
Fig. 5 shows that  
averaging with a Gaussian of width
$\Delta N^{1/2}=1.6$ ( corresponding to a 10\% spread in $N$ for 
$N=1000$) destroys only the basic SS, whereas the SSS remains visible.

Nowadays it is possible to manufacture probes with a large 
number of circular quantum dots, specifying the radius and the gate 
voltage with a 5\% accuracy \cite{weiss}. 
Changing the number of electrons in the dots by means of the gate 
voltage seems to be a possibility to measure the SSS of the
susceptibility. The imperfections in manufacturing
circular dots will have similar effects as discussed above for the 
metal clusters. The discussion of the 3D case can directly be applied to 
the 2D case.
Thus, the SSS pattern is expected to survive if the surface roughness
$\sigma <0.4/k_F$.

\begin{figure}[t]
\vskip-1cm
%\hspace{-4cm}
\hspace{-2cm}
%\mbox{\psfig{file=disk.ps,width=18cm}}
\mbox{\psfig{file=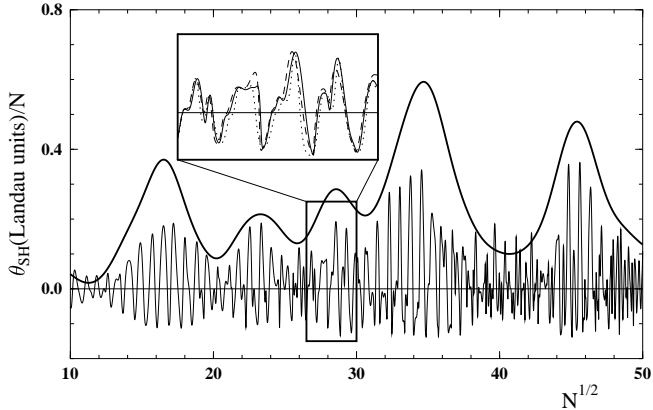,width=9.0cm}}
\vspace{-5cm}

\caption{
Shell contribution $\theta_{SH}$ to the magnetic susceptibility of
$N$ electrons in a circular well. The effective mass is $M=0.067M_e$, 
the well radius $R=500$ nm and the temperature $T=5\hbar^2/2MR^2 
\approx 0.13 K$~. 
The  line conventions are the same as in Fig.~1.
  Averaging over  $N^{1/2}$ is carried out with a Gaussian 
of width $\Delta N^{1/2}=1.6$. The non averaged values are divided 
by a factor of 10. }
\end{figure}

Fig. 6 demonstrates that, at variance with the 3D case,
 $F_{SH}(N)$ significantly differs from 
$\theta_{SH}(N)$.  For 2D case, the diameter orbit is 
not suppressed in $F_{SH}$. The  interference of the diameter, 
triangle and square results in slower basic and faster beat
oscillations as compared to $\theta_{SH}$, for which the 
diameter is missing.
Averaging with respect to $N^{1/2}$ filters out the slow
oscillation due to the interference between 
triangle and square, which mainly modulates  $F_{SH}(N)$.
The SS of the free energy should show up as a modulation of the 
capacitance of the dot. It is given by $d^2F/dN^2$, which also determines  
the abundances of heavy clusters \cite{hansen}.

\section{ Conclusions}
  The susceptibility of electrons confined in two or 
three dimensions by a spherical potential 
oscillates as function of their number. This shell structure is modulated 
by slow oscillations, the Super Shell Structure, which only develops at
sufficiently high temperatures.
% which originate from the  quantum interference of electrons on
%triangular and square periodic orbits.
Measurements that average out the shell structure may still reveal     
the Super Shell Structure.  
The free energy of electrons confined in three dimensions,
shows the analogous Super Shell pattern, which is  
observed  in the  abundances of metal clusters. 
However, for the two dimensional potential the  shell structure of the free 
energy  differs considerably from the one of the susceptibility.

\begin{figure}[t]
\vskip-1cm
\hspace{-2cm}
%\hspace{-4cm}
%\mbox{\psfig{file=shellcor.ps,width=18cm}}
\mbox{\psfig{file=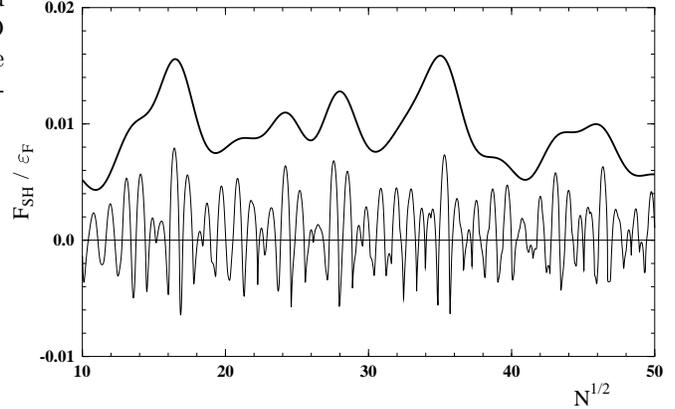,width=9.0cm}}
\vspace{-5cm}

\caption{
Shell contribution $F_{SH}$ to the free energy of
$N$ electrons in a circular well in units of
$\varepsilon_F=N\hbar^2/MR^2$. The calculation and the
 conventions are identical with Fig. 3}
\end{figure}

Discussions with  M.Brack and F.A.Ivanyuk and
financial support by INTAS
(93-0151) are gratefully acknowledged. 

%\end{document}
%\begin{thebibliograpy}

%\end{thebibliography}

\end{document}